\documentclass[reprint,showpacs,showkeys,preprintnumbers,amsmath,amssymb,aip,apl]{revtex4-1}

\usepackage{graphicx}
\usepackage{float}
\usepackage{color}

\begin{document}

\preprint{Rong Shan et al., NiYBi}

\title{A $p$-type Heusler compound:\\
       Growth, structure, and properties of epitaxial thin NiYBi films on MgO(100).}

\author{Rong~Shan}
\affiliation{Institut f{\"u}r Anorganische und Analytische Chemie,
             Johannes Gutenberg - Universit{\"a}t, 55099 Mainz, Germany.}
\affiliation{IBM Almaden Research Center
             San Jose, CA 95120, USA.}
                          
\author{Siham~Ouardi}
\affiliation{Max Planck Institute for Chemical Physics of Solids,
             01187 Dresden, Germany.}
             
\author{Gerhard~H.~Fecher}
\affiliation{Max Planck Institute for Chemical Physics of Solids,
             01187 Dresden, Germany.}

\author{Li~Gao}
\affiliation{IBM Almaden Research Center
             San Jose, CA 95120, USA.}
             
\author{Andrew~Kellock}
\affiliation{IBM Almaden Research Center
             San Jose, CA 95120, USA.}
            
\author{Andrei~Gloskowskij}
\affiliation{Institut f{\"u}r Anorganische und Analytische Chemie,
             Johannes Gutenberg - Universit{\"a}t, 55099 Mainz, Germany.}

\author{Carlos~E.~Vidal~Barbosa}
\affiliation{Max Planck Institute for Chemical Physics of Solids,
             01187 Dresden, Germany.}      

\author{Eiji~Ikenaga}
\affiliation{Japan Synchrotron Radiation Research Institute (JASRI), SPring-8, Hyogo 679-5198, Japan.}
        
\author{Claudia~Felser}
\affiliation{Max Planck Institute for Chemical Physics of Solids,
             01187 Dresden, Germany.}

\author{Stuart~S.~P.~Parkin}
\email{Stuart.Parkin@us.ibm.com}
\affiliation{IBM Almaden Research Center
             San Jose, CA 95120, USA.}

\date{\today}

\begin{abstract}

Epitaxial semiconducting NiYBi thin films were directly prepared on MgO(100) 
substrates by magnetron sputtering. The intensity ratio of the (200) and (400) 
diffraction peaks, $I(200) / I(400)=2.93$, was close to the theoretical value 
(3.03). The electronic structure of NiYBi was calculated using WIEN2k and a 
narrow indirect band gap of width 210~meV was found. The valence band spectra of 
the films obtained by linear dichroism in hard X-ray photoelectron spectroscopy 
exhibit clear structures that are in good agreement with the calculated band 
structure of NiYBi.

\end{abstract}



\maketitle


Heusler compounds of formula $XYZ$ and structure $C1_b$ are attractive 
candidates for electronic applications because they become semiconducting at a 
valence electron concentration of 6~$e^-$ per atom. $X$ and $Y$ stand for 
elements of the $3^{rd}$ to the $11^{th}$ group of the periodic table, and $Z$ 
represents an element from the $13^{th}$ to the $15^{th}$ group. The great 
diversity of Heusler compounds makes them versatile functional materials in 
different fields. For example, CoTiSb, NiTiSn, and similar compounds exhibit 
large Seebeck coefficients and are good candidates as thermoelectric 
materials~\cite{YLW08,OFB10}. Compounds with heavy elements (PtLuBi, PtLuSb, 
PtLaBi, etc.) have very recently been predicted to be promising topological insulators (TIs)\cite{CQK10}.
Because of the structural symmetry of Heusler TIs, the band gap around the Dirac point is closed. One feasible way to 
realize TI states may be the fabrication of quantum well 
structures using Heusler compounds. However, the lattice constants of almost all 
Heusler TIs are in the range from 6.4~{\AA} to 7.0~{\AA}. The epitaxial film 
growth of Heusler TIs also needs quite a high annealing temperature to ensure 
perfect film quality. The above two aspects hamper the use of conventional 
semiconductors as barrier layers, and thus Heusler semiconductors become better 
candidates for this purpose. Although the Heusler compounds are attracting broad 
interest, studies of semiconducting Heusler films are rare so far. In this work, 
semiconducting NiYBi films were produced and investigated. Such films are 
expected to be good candidates for nanostructured 
thermoelectrics~\cite{JMS11,XJO11}, as well as components of Heusler TI devices, because 
of their suitable lattice parameters and high melting points. Previously, thin films 
of NiHfSn~\cite{WCW10} or NiTiSn and NiZr$_{0.5}$Hf$_{0.5}$Sn~\cite{JMS11,XJO11} 
Heusler compounds with 1:1:1 stoichiometries and $C1_b$ structures were grown by 
sputtering using different buffer layers. Such films exhibit $n$-type 
conductivity, as do most Heusler compounds used as thermoelectric materials. In the 
present work, thin NiYBi films with $p$-type character were grown directly on 
MgO substrates.

The epitaxially grown films were investigated with respect to composition, 
crystalline structure, transport properties and electronic structure. Concerning 
the latter, hard X-ray photoelectron spectroscopy (HAXPES) is a powerful method 
for probing both chemical states and bulk electronic structures of thin films and 
buried layers in a non-destructive way~\cite{FBG08}. The combination of HAXPES 
with polarized radiation for excitation significantly extends its applicability. 
The use of linearly $s$- and $p$-polarized light in HAXPES enables the analysis 
of the symmetry of bulk electronic states~\cite{OFK11}. In the present study, the 
valence band electronic structures of NiYBi thin films were investigated using HAXPES and linear dichroism.


Samples with structures MgO(100)/NiYBi (20, 40, 80~nm) were prepared at 
different substrate temperatures from a composite target with a stoichiometry 
of Ni:Y:Bi = 1:1:1.2 by direct-current magnetron sputtering under an Ar atmosphere. The base pressure 
of the sputtering chamber was below $5\times 10^{-7}$~Pa. The films were capped 
by MgO(10~nm) layers to prevent oxidation. The compositions of the films were 
analyzed using Rutherford backscattering spectrometry and particle-induced 
X-ray emission. 

\begin{figure}
\centering
   \includegraphics[width=6.5cm]{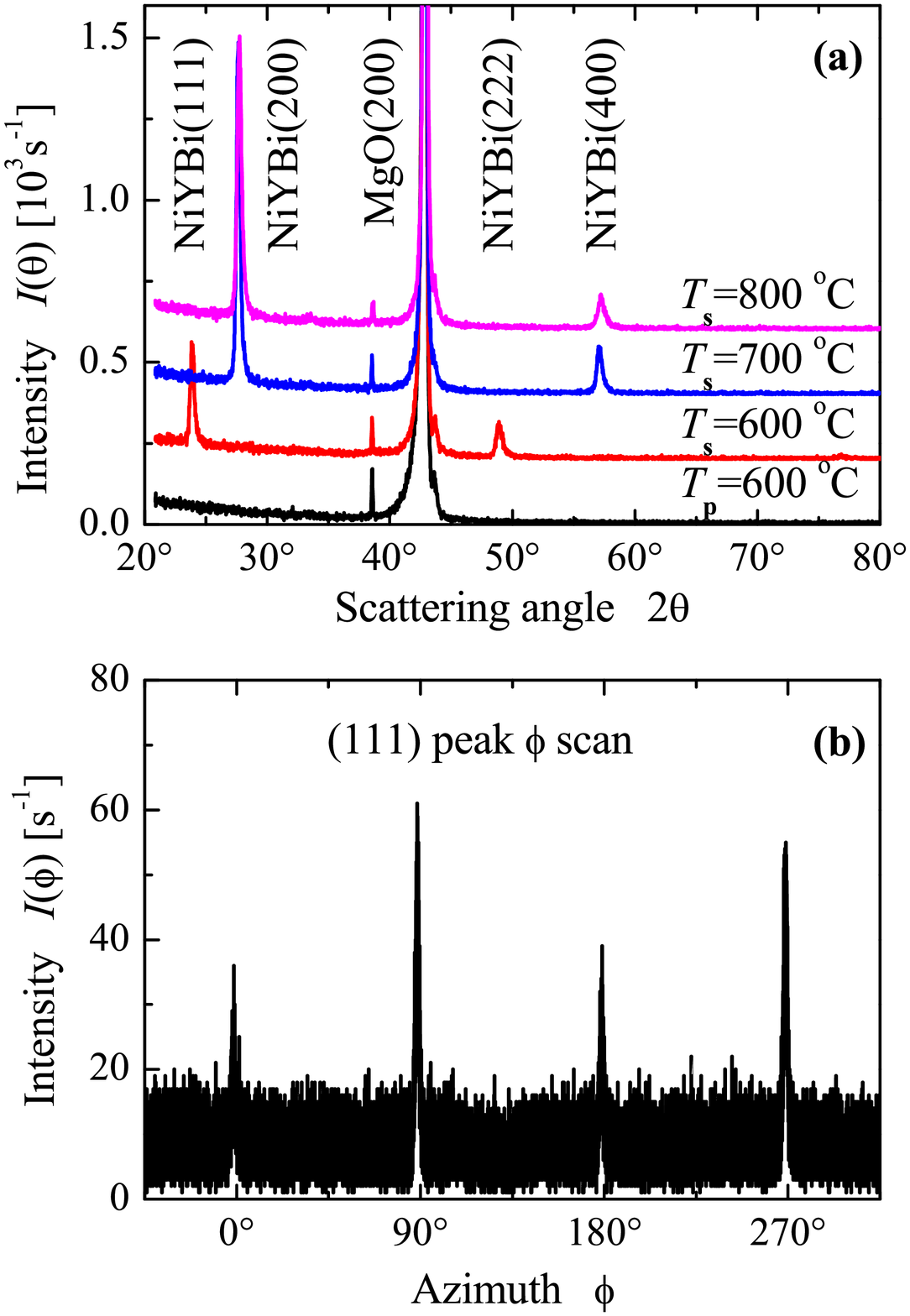}
   \caption{(Color online) X-ray diffraction of NiYBi thin films on MgO(100). \\
            (a) Polar scans of films prepared at different temperatures. 
                $T_p$ is the post-annealing temperature and 
                $T_s$ is the substrate temperature. 
            (b) NiYBi(111) azimuth scan.}
\label{fig:xrd}
\end{figure}

The Ni:Y:Bi ratio of the films depends on the growth temperature and was found 
to be about 26.7:38.6:34.7 at $600^\circ$C, 32.3:38.3:29.4 at $700^\circ$C, and 
32.7:38.6:28.7 at $800^\circ$C, with an uncertainty of $\pm1$\%. Obviously, Bi is 
slowly lost with increasing annealing temperature. Figure~\ref{fig:xrd} shows 
X-ray diffraction patterns of NiYBi thin films of thickness 40~nm on 
MgO(100) substrates. Here, $T_p$ means that the film was deposited at room 
temperature and post-annealed at temperature $T_p$; $T_s$ is the substrate 
temperature during film growth. The peak at around $38^\circ$ arises from the 
aluminum sample holder and corresponds to Al(111). There was no clear crystal 
structure formed when the films were prepared at room temperature, even if they were 
post-annealed at $600^\circ$C, as shown in Figure~\ref{fig:xrd}(a). The (111) 
orientation appears in the film deposited at $600^\circ$C and the films grow 
along the (100) orientation when $T_s$ is higher than $700^\circ$C. The lattice 
constant of NiYBi is around 6.41~{\AA}. The direct mismatch between NiYBi(100) 
and MgO(100) is very large ($\approx54$\%), but the mismatch (1.7\%) becomes 
acceptable if a NiYBi lattice can be built on one-and-a-half MgO lattice 
parameters with their (100) planes parallel to each other. To confirm this 
point, a (111) in-plane polar scan was used, as shown in 
Figure~\ref{fig:xrd}(b). The four-fold symmetry of the (111) reflection proved 
that {\it ``epitaxial''} NiYBi films were directly grown on the MgO(100) substrate. 
The intensity asymmetry arises from the rectangular shape of the 
sample. The $C1_b$ structure allows for three different structures, 
corresponding to Ni, Y, or Bi atoms at the Wyckoff position (4c) of the lattice. From 
these, the structure with Ni at (4c) has the lowest energy and thus is the most 
stable. Although the peak positions in the X-ray patterns are the same, the 
intensity ratios are different for the three structures. A peak ratio of 
$I(200)/I(400)\approx2.93$ was found for the film prepared at $700^\circ$C. This 
is close to the theoretical value (3.03), implying an ideal atomic arrangement 
in the lattice of the NiYBi film grown under this condition.

\begin{figure}
\centering
   \includegraphics[width=8.5cm]{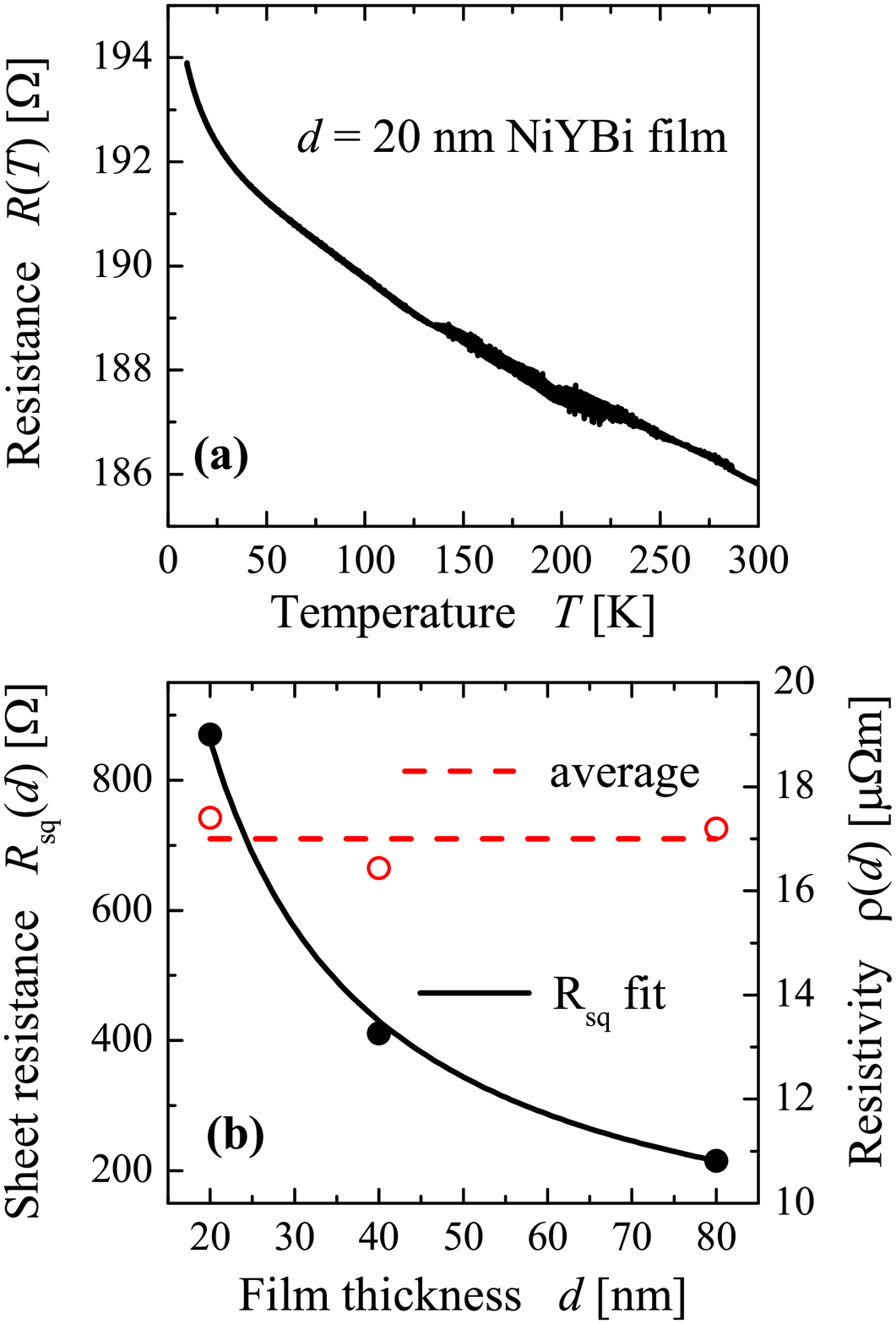}
   \caption{ (Color online) Electrical resistance of NiYBi thin films. \\
             (a) Temperature-dependent resistance of a 20~nm NiYBi film. 
             (b) Thickness dependence of sheet resistance (full symbols) 
                 and resistivity (open symbols).}
\label{fig:res}
\end{figure}

The temperature dependence of the resistance of a 20~nm NiYBi film prepared at 
$700^\circ$C is shown in Figure~\ref{fig:res}(a). The resistance decreases 
slowly with increasing temperature and suggests weak semiconducting behavior 
of the NiYBi films. This behavior was also observed in other half-Heusler 
semiconductors and is a typical characteristic of half-Heusler semiconductors as a result of 
heavy doping arising from composition deviations\cite{CTP91,OPR03}. The low 
resistivity of the film results from the same reason rather than from the width of 
the band gap, which can be significantly enhanced by carrier neutralization in the film 
through doping~\cite{OFB10}. The root-mean-square roughness of the film prepared 
at $700^\circ$C is about 2.8~nm. The maximum value between peak and valley is 
approximately 29.8~nm. In order to check the influence of roughness on 
resistivity in NiYBi films, three samples with thicknesses of 20, 40, and 80~nm 
were deposited at $700^\circ$C. Figure~\ref{fig:res}(b) shows the thickness 
dependence of the sheet resistance and resistivity. The fitted resistivity based 
on the sheet resistance (17.2~$\mu\Omega$m) is the same as the average 
resistivity (17~$\mu\Omega$m) measured from the three samples directly. This 
proves that the roughness hardly affects the resistance of the films.

\begin{figure}
\centering
   \includegraphics[width=8.5cm]{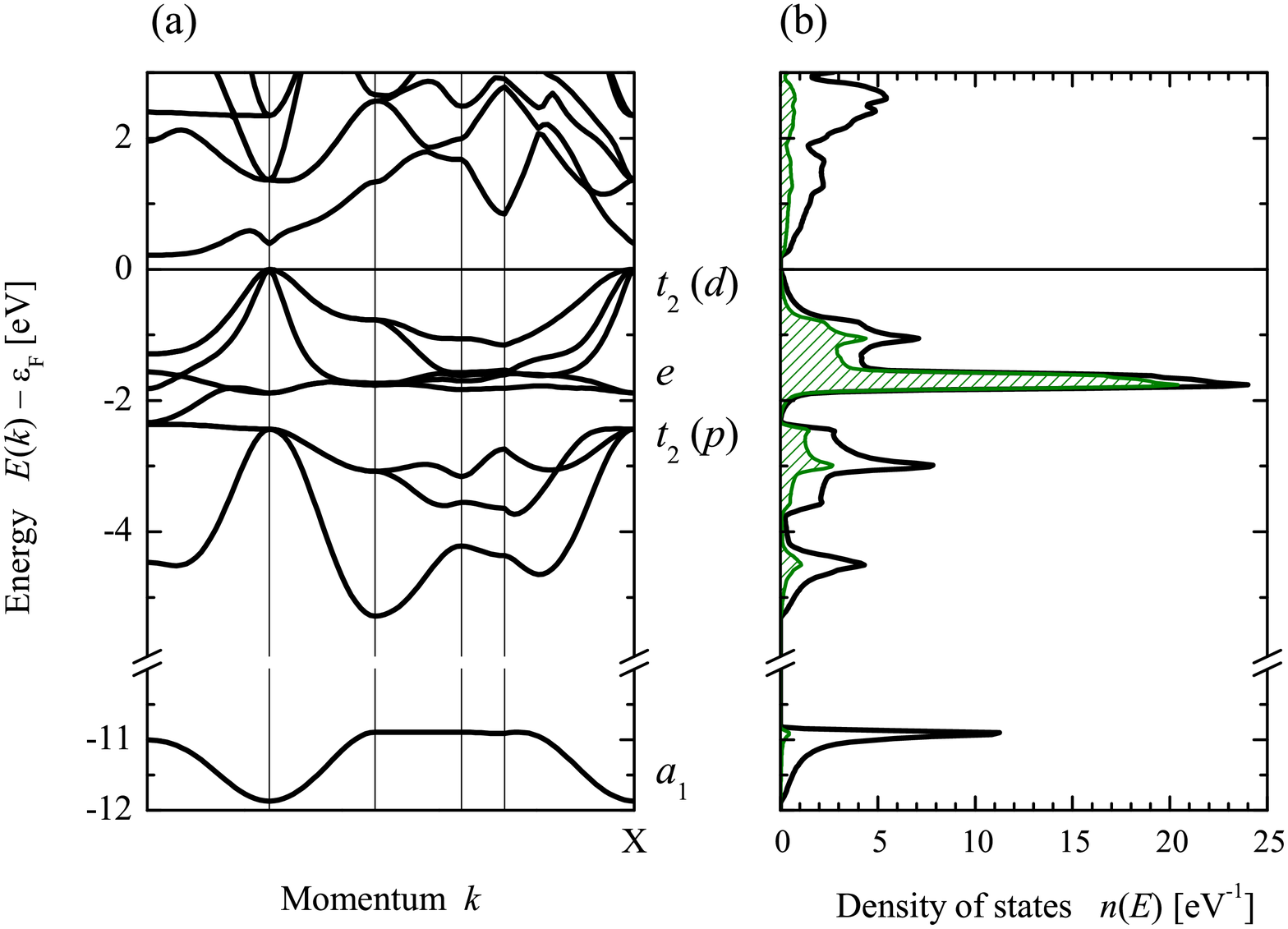}
   \caption{ (Color online) Electronic structure of NiYBi.\\
            (a) shows the band structure and (b) the density of states.
            The density localized at the Ni atoms is marked by the shaded area.
            The labels assign the irreducible representations of the states at the
            $\Gamma$ point.}
\label{fig:es}
\end{figure}

The electronic structure of NiYBi was calculated using {\sc Wien2k}~\cite{BSM01}, 
as described in detail in Reference~\cite{OFB10}. The result is shown in 
Figure~\ref{fig:es}. The compound turns out to be a narrow gap semiconductor. 
The indirect gap between $\Gamma$ and $X$ has a width of $\Delta E_{\rm 
gap}=215$~meV. The optical gap at $\Gamma$ is also small, with a width of only 
390~meV. The valence bands touching the Fermi energy have $t_2\:(d)$ character 
at $\Gamma$. The lower-lying states that cause a very high density of states at 
about -1.8~eV have $e$ character. The bottom of the upper part of the valence 
band is formed by states of $t_2\:(p)$ character at $\Gamma$. The $a_1$ states 
emerging from the Bi $s$ electrons are split off from the valence band and appear 
at about -11~eV.

\begin{figure}
\centering
   \includegraphics[width=6.5cm]{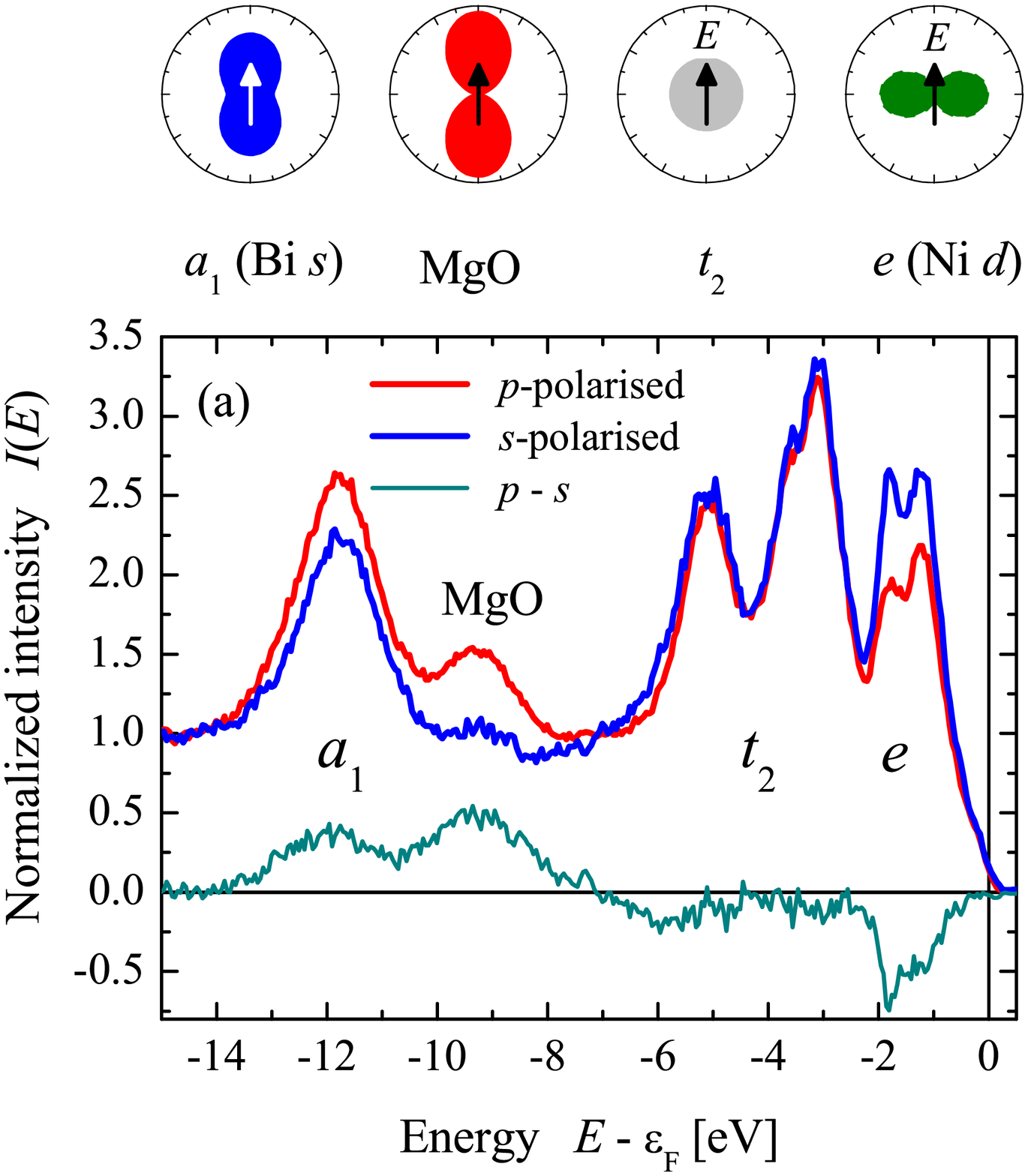}
   \caption{(Color online) Valence band spectra of NiYBi.\\
            The figure shows the polarization-dependent valence band spectra $I^p$, $I^s$, and the difference $I_{\rm LD}$.
            The spectra were obtained with a photon energy of about 8~keV.
            The emission characters of the different states are shown above the spectra
            in the upper part (note that the schematic intensities are not scaled by the cross-sections).
            Arrows indicate the direction of the electric field vector. }
\label{fig:haxpes}
\end{figure}

The HAXPES experiment was performed at BL47XU of Spring-8 (Japan). For details 
of the HAXPES experiment, see~\cite{OFK11,KSF11,XJO11}. The valence band spectra of 
NiYBi are presented in Figure~\ref{fig:haxpes}. The figure shows the valence band 
spectra excited by photons of energy about 8~keV with different polarizations and 
linear dichroism ($I_{\rm LD}=I^p-I^s$). The high intensity at about -13 to -10~eV
corresponds to excitation of the low-lying $a_1$ bands, as can be seen by 
comparison with the density of states (see Figure~\ref{fig:es}(b)). The intensity 
maximum centered at about -9~eV has no counterpart in the density of states of 
NiYBi. This arises from the MgO protective overlayer. Similar structures were 
previously also observed in the spectra from MgO-covered 
Co$_2$MnSi~\cite{FBG08}. The upper part of the valence spectra above -6~eV 
exhibits the typical structure of NiYBi valence bands with three major 
maxima. In particular, the topmost of the three maxima exhibits a pronounced 
splitting. The spectra exhibit a clear cut-off at the Fermi energy. This is 
typical of $p$-type materials, where the Fermi--Dirac distribution terminates the 
density of states slightly inside the valence band that is not fully occupied 
by electrons.

More differences in the intensities become obvious when the photon 
polarization is changed. In particular, for $l=0$ electrons, one expects lower intensities when 
using $s$-polarized photons. This is because the angular asymmetry parameter $\beta$ 
becomes 2 for $l=0$ states in the spherical approximation, resulting in 
a vanishing intensity for $s$-polarization. Similar to other Heusler 
compounds (compare with Reference~\cite{OFK11}), the intensity of the $a_1$ states 
does not vanish but is only reduced. The effect of the polarization is more 
pronounced at the MgO states (note the lower intensity but larger difference 
between the spectra in this energy region). This points to a $\beta$ parameter 
closer to 2 for these states. The relation between the linear dichroism 
(asymmetry $A_{\rm LD}=\frac{I_{\rm LD}}{I^p+I^s}$) and the angular asymmetry 
parameter $\beta$ is given by

\begin{equation}
	\beta = \frac{4A_{\rm LD}}{1+A_{\rm LD}}
\end{equation}

for complete linear polarization of the photons. The linear dichroism 
asymmetries of the MgO and NiYBi $a_1$ states are (after subtraction of 
linear backgrounds) about 92\% and 22\%, resulting in angular asymmetry parameters 
of $\beta_{\rm MgO}\approx1.9$ and $\beta_{a_1}\approx0.7$, respectively. 
Assuming a slightly incomplete $s$-polarization as a result of a small misalignment of 
the phase retarder, $\beta_{\rm MgO}$ might come close to the expected value of 
2, but even in that case, the $\beta$ parameter of the NiYBi $a_1$ state with 
cubic symmetry stays clearly below 1 and thus has a large deviation from 
spherically symmetric, atomic-type states with pure $l=0$ ($s$) character. The 
emission characteristics resulting from the different $\beta$ parameters are 
shown in the upper part of Figure~\ref{fig:haxpes}.

Most parts of the valence band above -6~eV do not differ much when the 
polarization is changed, that is, the emission is rather spherically symmetric. The most 
distinct difference appears at -1.75~eV, where the intensity for $s$-
polarized light is higher than that of $p$-polarized light. This is a result of 
the different characters of the $t_2$ and $e$ states. Here, this points to 
strong localization of the $e$ states at Ni, as is seen from the electronic 
structure in Figure~\ref{fig:es}.


In conclusion, $p$-type semiconducting epitaxial NiYBi films were prepared 
directly on MgO(100) substrates. X-ray diffraction and HAXPES confirmed the good quality of 
the films in terms of crystalline and electronic structures, respectively. 
This material could therefore be a promising buffer layer or barrier in 
TIs, spintronic devices, and nanostructured thermoelectrics 
based on Heusler compounds, although the composition of the films may need to be 
further tuned for the specific requirements of these potential application.

\begin{acknowledgments}

Financial support by the DFG-JST (project P~1.3-A in research unit FOR 1464 {\it 
ASPIMATT}) is gratefully acknowledged. HAXPES was performed at BL47XU of SPring-8 
with approval of JASRI (Proposal No.~2011B1566).

\end{acknowledgments}

%

\end{document}